\documentclass[twocolumn,showpacs,preprintnumbers,amsmath,amssymb]{revtex4}

\usepackage{graphicx}
\usepackage{dcolumn}
\usepackage{bm}

\begin{document}


\title{Experimentally accessible reentrant phase transitions
in double-well optical lattices}
\author{Ippei Danshita$^{1,2}$}
\author{Carlos A. R. S\'a de Melo$^{1,3}$}
\author{Charles W. Clark$^1$}
\affiliation{
{$^1$Joint Quantum Institute, National Institute of Standards and Technology, and University of Maryland, Gaithersburg, Maryland 20899, USA}
\\
{$^2$Department of Physics, Waseda University, Shinjuku-ku, Tokyo 169-8555, Japan}
\\
{$^3$School of Physics, Georgia Institute of Technology, Atlanta, Georgia 30332, USA}
}

\date{\today}

\begin{abstract}
We study the quantum phases of bosons confined in a combined potential
of a one-dimensional double-well optical lattice and a parabolic trap. 
We apply the time-evolving block decimation method to the corresponding 
two-legged Bose-Hubbard model. 
In the absence of a parabolic trap, the system of bosons in the double-well 
optical lattice exhibits a reentrant quantum phase transition between 
Mott insulator and superfluid phases at unit filling 
as the tilt of the double-wells is increased.
We show that the reentrant phase transition occurs also in the presence of
a parabolic trap and suggest that it can be detected in experiments
by measuring the matter-wave interference pattern.

\end{abstract}

\maketitle
Recently, the system of ultracold bosonic atoms in a double-well optical 
lattice (DWOL) has been used to study quantum 
information processing and quantum many-body systems
~\cite{rf:jenni1,rf:jenni2,rf:marco2,rf:foelling}.
A DWOL is constructed from two-color two-dimensional (2D) lattices in the 
horizontal $xy$ plane and another lattice in the vertical $z$ direction, 
resulting in a 3D lattice whose unit cell is a double well.
By exploiting precise control of the properties of the double wells,
recent experiments have realized
an atom interferometer~\cite{rf:jenni2},
 two-atom exchange oscillations, which were used to create
a quantum SWAP gate~\cite{rf:marco2}, 
and correlated tunneling of two interacting atoms~\cite{rf:foelling}.

In single-well optical lattices, the transition from a superfluid (SF) phase 
to a Mott insulator (MI) phase has been observed by adiabatically increasing 
the lattice depth~\cite{rf:greiner,rf:ian}.
In previous DWOL experiments
all the double wells were independent of each other due to the large 
potential barrier between them, and the system was always in an insulating 
phase.
However, the potential barrier could be lowered to induce 
a transition to a SF phase~\cite{rf:jenni1}.
In particular, one can investigate the SF-MI transition in a two-legged 
ladder 1D DWOL by tuning the lattice depth in the vertical direction.
While in single-well lattices the SF-MI transition is determined
by the ratio, $U/t$, of the on-site interaction energy to the 
hopping energy~\cite{rf:greiner,rf:ian},
in DWOL the SF-MI transition is governed by the competition between
the on-site interaction and the tilt (energy offset) of the double wells. 

In this paper, we study the ground state properties of bosons in 
a two-legged ladder lattice with a parabolic confining potential.
This case is more representative of current experiments than is 
an unconfined, homogeneous lattice.
Although mean-field-type methods describe the SF-MI transition of bosons in
optical lattices very well in 3D~\cite{rf:oosten}, 
they fail to describe quantitatively the two-legged ladder system 
due to strong quantum fluctuations~\cite{rf:danshita}.
Hence, we use the time-evolving block decimation (TEBD) method~\cite
{rf:vidal1,rf:vidal2}, which is one of the best methods currently available
to study strongly correlated 1D quantum systems.
We show here that a reentrant phase transition (MI-SF-MI) occurs 
as the tilt of the double wells is changed for a fixed parabolic confinement.
Moreover, a distinctive signature of the reentrant phase transition 
appears in the momentum distribution, which is directly related to
matter-wave interference patterns observed in experiments.

%
\begin{figure}[tb]
\includegraphics[scale=0.20]{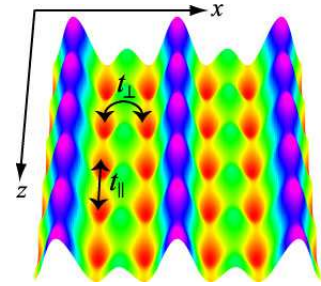}
\caption{\label{fig:ladder}
(color online)
Schematic double-well lattice potential in the $xz$ plane for $\alpha = 0.2$, $\theta = 0$,
showing the existence of isolated two-legged ladders.
}
\end{figure}
%

We consider a Bose gas at zero temperature confined 
in a combined potential of a DWOL and a parabolic trap.
A DWOL potential in Refs.~\cite{rf:jenni1,rf:jenni2,rf:marco2}
 is given by
\begin{eqnarray}
\!\!\!\!
V(\boldsymbol{r})\!\!&=&\!\! 
           V_{\rm in}\left({\rm sin}^2(kx)+{\rm sin}^2(ky)-2\right) 
\nonumber\\ 
           & &\!\!\!\!\!\!\!\!\!
           -V_{\rm out}\left({\rm sin}(kx-\theta)+{\rm cos}(ky)\right)^2
           +V_{\rm z}\,{\rm sin}^2(kz),\label{eq:DWOL}
\end{eqnarray}
where $V_{\rm in}$ is the depth of the 
``in-plane" lattice with period of $d\equiv \pi/k$, whose light polarization is in the $xy$ plane,
$V_{\rm out}$ the depth of the ``out-of-plane" lattice  with 
$2d$-periodicity~\cite{rf:jenni1}, and 
$V_{\rm z}$ is the depth of the lattice in the vertical $z$ direction.
When the ratio $\alpha\equiv V_{\rm out}/V_{\rm in}$ satisfies $0<\alpha<0.5$,
the unit cell of the potential Eq.~(\ref{eq:DWOL}) is a double-well.
Control of the ratio $\alpha$ and the relative phase $\theta$ provides
the flexibility to adjust the double-well parameters: the barrier height and
the tilt.
When $\theta=0$, the double wells are symmetric (no tilt).
Increasing $\alpha$ from zero enlarges the barrier height $V_{\rm high}$ 
between double wells in the $xy$ plane and reduces the barrier height 
$V_{\rm low}$ between two wells within a double well.
Keeping $V_{\rm high}-V_{\rm low}\gg E_{\rm R}$ and tuning $V_{\rm z}$, 
one can create an ensemble of 1D DWOLs with a two-legged ladder 
structure, where all the two-legged ladders are decoupled 
from each other as shown in Fig.~\ref{fig:ladder}. 
Here $E_{\rm R}=\frac{\hbar^2 k^2}{2m}$ is the recoil energy, 
where $m$ is the boson mass.

At zero temperature the many-body quantum state of 
interacting bosons in an optical lattice is well described by 
the Bose-Hubbard model when the lattice is so deep that only the lowest energy
level of each lattice site is occupied and tunneling occurs only between 
nearest-neighboring sites~\cite{rf:fisher,rf:jaksch}.
To study bosons in 1D DWOLs, we use the two-legged 
Bose-Hubbard Hamiltonian (BHH)
\begin{eqnarray}
\label{eq:hamil}
H \!&=&\! \sum_{j}\biggl
\{\sum_{\eta\in\{L,R\}}\Bigl(
\frac{U}{2} \hat{n}_{j,\eta} (\hat{n}_{j,\eta}-1)
+(\Omega j^2-\mu)\hat{n}_{j,\eta}
\nonumber\\
&&\!\!
-t_{\parallel}(a^{\dagger}_{j+1,\eta} a_{j,\eta}\!+\!{\rm h.c.})
\Bigr)
-t_{\perp}\!( a^{\dagger}_{j,R} a_{j,L}\!+\!{\rm h.c.})
\nonumber\\
&&+\frac{\lambda}{2}(\hat{n}_{j,L}-\hat{n}_{j,R})
\biggr\}.\label{eq:hamiltonian}
\end{eqnarray}
$a^{\dagger}_{j,\eta}$ creates a boson at the lowest level localized 
on the left (right) of the {\it j}-th double-well for $\eta=L(R)$, and
$\hat{n}_{j,\eta}=a^{\dagger}_{j,\eta}a_{j,\eta}$ is the number operator.
The lattice parameters can be related to the DWOL depths $V_{\rm in}$, $V_{\rm out}$, 
and $V_z$ and the recoil energy $E_{\rm R}$ by assuming that 
$\theta \ll 1$ and that each well is sufficiently deep to approximate
the Wannier function by a Gaussian. 
In this case, the on-site interaction is $ U  \sim E_{\rm R} (8\pi)^{1/2}
\frac{a_{\rm s}(1 - 2\alpha)^{1/4}}{d( 1 - \alpha )^{1/2} } s_{\rm z}^{1/4}s^{1/2} $;
the intrachain hopping is 
$t_{\parallel} \sim
E_{\rm R} \left( \frac{\pi^2}{4} - 1 \right) s_{\rm z} \, e^{-\pi^2\sqrt{s_{\rm z}}/4}$;
the interchain hopping is 
$t_{\perp} \sim E_{\rm R} \left( \sqrt{ \frac{1 - 2 \alpha}{1 -\alpha} } 
f(\alpha) - 1 + 2\alpha \right) s\, e^{ -f(\alpha)\sqrt{s} }$;
the tilt (energy offset) of the double wells is 
$\lambda \sim E_{\rm R} \frac{4\alpha\sqrt{1-2\alpha}}{(1-\alpha)^2} s \theta$.
Here, $\alpha = V_{\rm out}/ V_{\rm in}$, $s=V_{\rm in}/E_{\rm R}$,
$s_{\rm z}=V_{\rm z}/E_{\rm R}$, $a_{\rm s}$ is the s-wave scattering length,
and 
$
f(\alpha)=\sqrt{\frac{1-2\alpha}{1-\alpha}}{\rm arccos}^2 \left(\frac{\alpha}{1-\alpha}\right).
$
$\Omega$ is the curvature of the parabolic trap,
and $\mu$ is the chemical potential, 
which controls the total number of bosons $N$. 
Notice that $U$, $t_{\parallel}$, and $t_{\perp}$ are approximately independent of the tilt
as long as $\theta\ll 1$.

\begin{figure}[tb]
\includegraphics[scale=0.39]{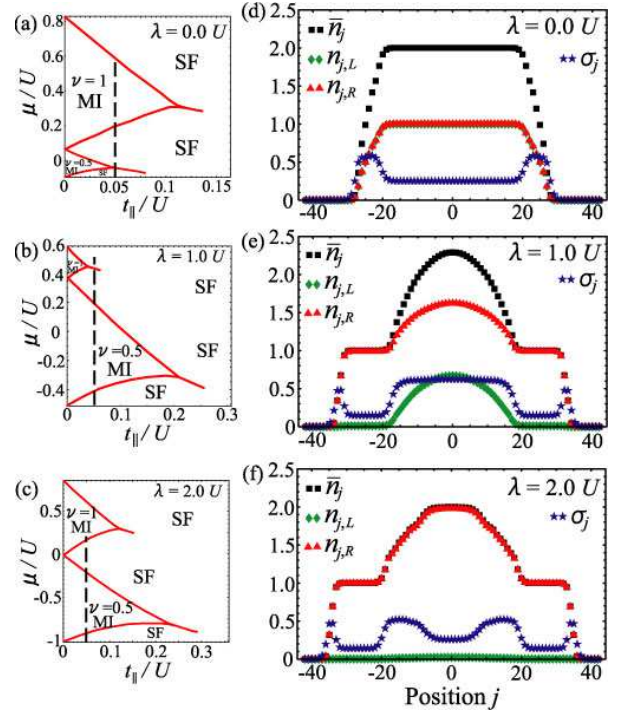}
\caption{\label{fig:densities}
(color online)
Results at three values of $\lambda/U$: $0$ (a) and (d), $1$ (b) and (e),
and  $2$ (c) and (f).
In the left three figures, we show the phase diagrams of the homogeneous
system in the ($\mu/U,t_{\parallel}/U$) plane for $t_{\perp}=0.1 U$, 
which have been calculated by means of the infinite-size version of 
the TEBD method.
In the right three figures, the local number of bosons
at the $j$-th double-well $\bar{n}_j$ (squares),
that at the $j$-th left well (diamonds),
that at the $j$-th right well (triangles), 
and the fluctuation $\sigma_j$ of $\bar{n}_j$ (stars) are shown.
The dashed lines in the phase diagrams represent the values of 
the local chemical potential and the intrachain hopping realized in
the right figures.
The upper (lower) end of the dashed lines corresponds to the center (edges) of 
the trapped gas.
}
\end{figure}

We now describe briefly the zero-temperature phase diagrams of the two-legged 
Bose-Hubbard model with no confining potential 
($\Omega = 0$)~\cite{rf:danshita,rf:buonsante,rf:donohue}.
In Figs.~\ref{fig:densities}(a)-(c), we show the phase diagrams in the 
$(\mu/U,t_{\parallel}/U)$ plane for $t_{\perp}/U=0.1$.
There we see the usual MI phase with unit filling, $\nu=1$ 
(corresponding to two atoms per double well),  and also 
one with half filling, $\nu=1/2$~\cite{rf:danshita,rf:buonsante}.
The SF-MI phase boundary in the ($\mu/U,t_{\parallel}/U$) plane significantly
depends on $t_{\perp}/U$~\cite{rf:donohue} and $\lambda/U$~\cite{rf:danshita}.
In particular, the unit-filling MI phase depends
 nonmonotonically upon $\lambda/U$.
As seen in the upper lobes of the phase diagrams, the $\nu=1$ MI region 
shrinks initially as $\lambda/U$ is increased from zero.
The size of the MI region is minimized at $\lambda =U$, where the local states
$|1,1\rangle$ and $|0,2\rangle$ are nearly degenerate; 
$|n_L,n_R\rangle$ designates the local Fock state with $n_L (n_R)$ bosons on 
the left (right) of a double well.
Due to this degeneracy, coherence is developed within each double well and 
the system favors the SF phase. 
As $\lambda/U$ is increased further, the MI region grows again.
This nonmonotonic behavior leads to a reentrant phase transition from 
a MI to a SF and again to a MI, induced by increasing $\lambda/U$.

Since we are mainly interested in the reentrant phase transition,
we investigate how the ground state properties change 
with increasing $\lambda/U$ for the following fixed values of the 
other parameters; $N=98$, $\Omega/U=0.001$, 
$t_{\perp}/U=0.1$, and $t_{\parallel}/U=0.05$.
This value of $t_{\parallel}/U$ is sufficiently small 
such that the MI domain with $\nu=1$ is present 
when $\lambda=0$ or $\lambda \gg U$, while it is so large 
that the $\nu =1$ MI domain is absent when $\lambda = U$.
The number of double wells is chosen as $L=85$.
This value of $L$ is so large that the bosons do not see 
the edge of the system.

To analyze the two-legged BHH~(\ref{eq:hamiltonian}), we use 
the finite-size version of the TEBD method~\cite{rf:vidal1}, which provides 
a precise ground state for 1D quantum lattice systems 
via propagation in imaginary time.
While the maximum number of bosons per site is $N_{\rm max}=\infty$, 
convergence is already achieved in our numerical calculations 
when $N_{\rm max}=5$.
We calculate the ground states of the Hamiltonian~(\ref{eq:hamiltonian})
with $\chi=80$, where $\chi$ is the size of the basis set
retained in the TEBD procedure~\cite{rf:vidal1}.

\begin{figure}[tb]
\includegraphics[scale=0.3]{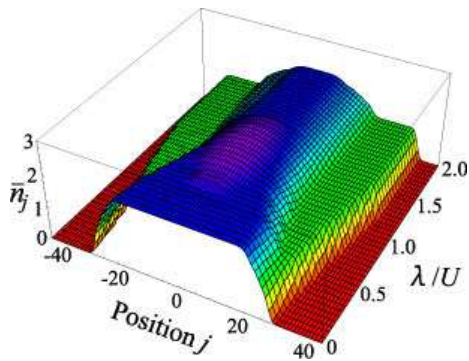}
\caption{\label{fig:densityoftilt}
(color online) Density profile $\bar{n}_j$ as a function of $\lambda/U$.
All the other parameters are fixed as $N=98$, $L=85$, $t_{\parallel}/U=0.05$,
$t_{\perp}/U=0.1$, and $\Omega /U=0.001$.
}
\end{figure}

In Fig.~\ref{fig:densityoftilt}, we show the density profile
%
$
\bar{n}_j = n_{j,L}+n_{j,R} = \sum_{\eta} \langle \hat{n}_{j,\eta}\rangle,
$
%
namely the local number of bosons in the $j$-th double well
as a function of $\lambda/U$. 
For $\lambda = 0$, the system is essentially a plateau with 
unit filling ($\bar{n}_j = 2$),
with small SF regions with incommensurate filling at the edges.
This plateau indicates the presence of an incompressible 
MI domain~\cite{rf:ian,rf:kollath,rf:kashurnikov,rf:batrouni,rf:wessel}.
As $\lambda/U$ is increased, the unit-filling plateau starts to melt 
at a certain point and a new MI plateau with half filling 
($\bar{n}_j=1$) emerges.
In the vicinity of $\lambda=U$, the unit-filling plateau has been completely
melted away and the density profile for $\bar{n}_j>1$ is smooth, 
with the reflected parabolic shape of the confining potential that
is characteristic of SF phases in the regime 
where the local density approximation is valid~\cite{rf:kollath}.
As $\lambda/U$ is increased further, the unit-filling plateau forms again
at the center of the system.
Thus, as in the homogeneous case, the reentrant phase transition is caused 
by increasing the tilt parameter
in the presence of a parabolic potential~\cite{rf:footnote}.

For a better understanding of the reentrant phase transition, three slices 
from Fig.~\ref{fig:densityoftilt} are shown in Figs.~\ref{fig:densities}(d)-(f)
together with $n_{j,L}$, $n_{j,R}$, and the fluctuation of $\bar{n}_j$,
%
$
\sigma_j = \sqrt{\langle \hat{n}_j^2 \rangle-\bar{n}_j^2},
$
%
which is small in the MI regions and relatively large in the SF regions.
One can roughly interpret the spatial dependence of $\bar{n}_j$ by a local
density approximation.
Introducing the effective local chemical potential 
$\mu^{\rm eff}_j = \mu - \Omega j^2$,
$\bar{n}_j$ is approximated by $\tilde{n}(\mu^{\rm eff}_j)$, where
$\tilde{n}(\mu)$ is the number of bosons per double well for uniform systems.
Scanning the effective chemical potential along the dashed lines in 
the phase diagrams of Figs.~\ref{fig:densities}(a)-(c), the behavior of
the local density in Figs.~\ref{fig:densities}(d)-(f) can be explained.
When $\lambda = U$, for example, the upper end of the dashed line in 
Figs.~\ref{fig:densities}(b) is located in a SF phase with $\tilde{n}>2$
as realized at the center of the trapped gas in Fig.~\ref{fig:densities}(e).
As $\mu$ is decreased from the upper end, the dashed line enters the MI phase
with half filling, corresponding to the half-filling plateau of
the trapped gas.
As $\mu$ is decreased further to the lower end, the dashed line exits to 
a SF phase with $\tilde{n}<1$, which is also realized at the boundary of
the trapped gas.
Notice that the local density approximation breaks down in the vicinity of 
the interfaces between the commensurate and incommensurate regions, 
where the singularities due to quantum critical behavior observed 
in uniform systems are removed~\cite{rf:batrouni,rf:wessel}.

In Figs.~\ref{fig:densities}(d)-(f), the population of bosons 
in the left wells shifts gradually to the right wells 
as $\lambda/U$ is increased.
At $\lambda = U$, almost all the bosons occupy the right wells, 
namely $n_{j,L}\simeq 0$, for $\bar{n}_{j}\le 1$.
In contrast, a considerable amount of bosons remains in the left well 
as $n_{j,L}\simeq \frac{\bar{n}_j-1}{2}$ for $\bar{n}_{j}>1$, since the local state 
with unit-filling is approximately given by 
$(|1,1\rangle+|0,2\rangle)/\sqrt{2}$ 
due to the competition between the onsite interaction and the tilt.
At $\lambda = 2U$, the population of bosons in the left well almost vanishes, namely $n_{j,L}\simeq 0$, in the entire system.

\begin{figure}[tb]
\includegraphics[scale=0.3]{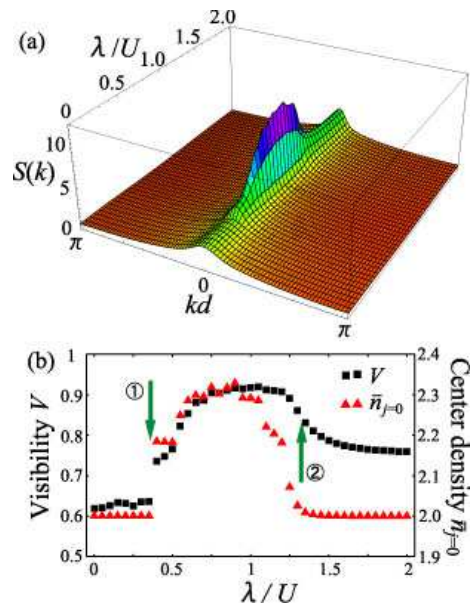}
\caption{\label{fig:Qmom}
(color online)
(a) Quasi-momentum distribution $S(k)$ as a function of $\lambda/U$.
(b) Visibility $V$ (squares) and center density $\bar{n}_{j=0}$ (triangles)
as functions of $\lambda/U$.
}
\end{figure}

Next, we calculate the quasi-momentum distribution
%
$
S(k)=L^{-1}\sum_{j,l}\sum_{\eta}e^{ik(j-l)d}
\langle a^{\dagger}_{j,\eta}a_{l,\eta}\rangle,
$
%
where $d$ is the lattice spacing.
Since the true-momentum distribution is expressed as the product of $|w(k)|^2$
and $S(k)$~\cite{rf:kashurnikov}, where $w(k)$ is the Fourier transform 
of the Wannier function in the lowest Bloch band, the quasi-momentum 
distribution can be extracted by dividing the true-momentum distribution 
observed in experiments by $|w(k)|^2$~\cite{rf:ian}.
$S(k)$ is shown as a function of $\lambda/U$ 
in Fig.~\ref{fig:Qmom}(a), where the $k=0$ peak is 
the sharpest in the vicinity of $\lambda = U$.
This means the intrachain coherence increases as 
the SF region around the trap center emerges.

To quantify the sharpness of the peak in $S(k)$,
we calculate the visibility of interference patterns defined as
%
$V=(S_{\rm max}-S_{\rm min})/(S_{\rm max}+S_{\rm min})$~\cite{rf:gerbier,
rf:sengupta}.
%
In Fig.~\ref{fig:Qmom}(b), we show $V$ and the center density $\bar{n}_{j=0}$
as a function of $\lambda/U$.
As indicated by the first arrow, the visibility 
exhibits a sudden jump at the point where the unit-filling MI plateau starts
to melt.
In the vicinity of $\lambda = U$, where the unit-filling MI plateau does not
exist, the visibility is distinctively large compared to the regions where
the unit-filling MI plateau is present.
As $\lambda/U$ is increased further, the visibility decreases the most rapidly
at the point indicated by the second arrow, where the $\nu =1$ MI plateau
forms again.
Notice that the change of the visibility at the second arrow is not as
distinctive as the jump at the first arrow because the $\nu =1$ Mott plateau 
at $\lambda =2U$ is small compared to that at $\lambda =0$.
This tendency of the visibility suggests that the reentrant phase transition 
can be characterized in experiments through the observation of matter-wave 
interference patterns.

\begin{figure}[tb]
\includegraphics[scale=0.27]{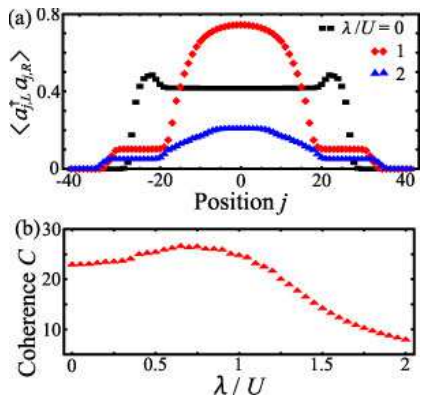}
\caption{\label{fig:coherence}
(color online)
(a) Local coherence between the left and right wells $\langle a^{\dagger}_{j,L}
a_{j,R} \rangle$ for $\lambda/U=0$ (squares), 1 (diamonds), and 2 (triangles).
(b) Total coherence $C$ versus $\lambda/U$.
}
\end{figure}

Another key observable in DWOLs is the coherence 
between the left and right wells
%
$
C 
  = \sum_j \langle a^{\dagger}_{j,L} a_{j,R} \rangle.
$
%
The coherence $C$ can be extracted from the double-slit diffraction pattern,
which has been observed in the time-of-flight images in the horizontal $xy$
plane~\cite{rf:jenni1,rf:jenni2,rf:marco2}.
The larger $C$ is, the clearer the diffraction pattern is.

We calculate first the local coherence $\langle a^{\dagger}_{j,L} a_{j,R} 
\rangle$ as shown in Fig.~\ref{fig:coherence}(a).
At $\lambda=0$, since the local state with half-filling is well approximated by
the bonding state, the SF regions at the boundary of the trapped gas have large
local coherence.
As $\lambda/U$ is increased, the region with $\bar{n}_j\le 1$ immediately 
loses the local coherence.
In contrast, in the region with $\bar{n}_j > 1$
the local coherence is pronouncedly large at $\lambda = U$.
The development of local coherence is due to the degeneracy of $|1,1\rangle$ and $|0,2\rangle$ states,
and it drives the system into the SF phase.
As $\lambda/U$ is increased further, the entire system tends to lose
coherence.
In Fig.~\ref{fig:coherence}(b), the total coherence $C$ is shown as a function 
of $\lambda/U$.
While the total coherence is significantly reduced when $\lambda$ is increased 
from $U$, it hardly changes in the region $\lambda<U$.
This happens because the gain of local coherence for $\bar{n}_j>1$ and the
loss for $\bar{n}_j\le 1$ almost cancel each other out.
Consequently, the double-slit diffraction pattern can not capture the 
signature of the reentrant phase transition.

In conclusion, we have studied the SF-MI transition of parabolically 
trapped Bose gases in a 1D double-well optical lattice by using the 
time-evolving block decimation (TEBD) method.
We have calculated the density profile as a function of the tilt of the double
wells and shown that a reentrant phase transition between MI and SF phases 
occurs in the presence of a parabolic confinement.
We have calculated also the quasi-momentum distribution and found that the 
matter-wave interference pattern, which is one of the most common observables
in experiments with cold atoms, contains sufficient information to characterize
the reentrant phase transition. 
We would like to emphasize that, unlike the results of mean-field theories,
our results based on the TEBD method are quantitative, thus we expect 
the reentrant phase transition to be observed in future experiments near 
the parameter values discussed in this paper.

We thank Jamie Williams and Trey Porto for stimulating discussions.
I. D. also thanks Gabriele De Chiara for helpful discussions regarding the TEBD algorithm.
I. D. acknowledges support from a Grant-in-Aid from JSPS, and C. SdM thanks NSF (DMR - 0709584)
for support.


\begin{thebibliography}{99}
\bibitem{rf:footnote}
Strictly speaking, because of the lack of critical behavior, the transition
between MI and SF phases in parabolically confined systems is not a true 
quantum phase transition~\cite{rf:batrouni,rf:wessel}.
\end{thebibliography}
\end{document}